\documentclass[a4paper,11pt]{article}
\usepackage{jcappub} % for details on the use of the package, please
\usepackage{natbib}
\usepackage[T1]{fontenc} % if needed
\usepackage{lineno}

\title{New X-ray bound on density of primordial black holes}

\author[a]{Yoshiyuki Inoue}
\author[b,c]{Alexander Kusenko}

\affiliation[a]{Institute of Space and Astronautical Science JAXA, 3-1-1 Yoshinodai, Chuo-ku, Sagamihara, Kanagawa 252-5210, Japan}
\affiliation[b]{Department of Physics and Astronomy, University of California, Los Angeles, CA 90095-1547, USA}
\affiliation[c]{Kavli IPMU (WPI), University of Tokyo, Kashiwa, Chiba 277-8568, Japan}

\emailAdd{yinoue@astro.isas.jaxa.jp}
\emailAdd{kusenko@ucla.edu}

\abstract{
We set a new upper limit on the abundance of primordial black holes (PBH) based on existing X-ray data.  PBH interactions with interstellar medium should result in significant fluxes of X-ray photons, which would contribute to the observed number density of  compact X-ray objects in galaxies. The data constrain PBH number density in the mass range from a few $ M_\odot$ to $2\times 10^7 M_\odot$.  PBH density needed to account for the origin of black holes detected by LIGO is marginally allowed. 
}

\begin{document}

%\linenumbers
\maketitle
\flushbottom

\section{Introduction}
%\tableofcontents
 Primordial black holes (PBH) could form in the early universe in a variety of plausible scenarios~\cite{Zeldovich:1967,Hawking:1971ei,car74,GarciaBellido:1996qt,Khlopov:2008qy,car10,Frampton:2010sw,Kawasaki:2012kn,kaw16,car16,Cotner:2016cvr,inomata16,Inomata:2017okj,Georg:2017mqk,Domcke:2017fix,Ezquiaga:2017fvi,Cotner:2017tir}.  They can account for all dark matter (DM) in a narrow mass window around $10^{20}$g, but even if PBH make up only a small fraction of DM, they can play a role in $r$-process nucleosynthesis~\cite{ful17}, provide seeds for supermassive black holes~\cite{Bean:2002kx,Kawasaki:2012kn}, or contribute to gravitational wave signals~\cite{Nakamura:1997sm,Clesse:2015wea,bir16,Garcia-Bellido:2017qal,Garcia-Bellido:2017aan}.

%Various limits on PBH abundance exist in the literature~\cite{car16}, but some of them have been questioned~\cite{bir16}, and there is a strong interest in finding additional reliable constraints. 

% Gravitational collapse of the Hubble patch in the early universe would form primordial black holes \citep[PBHs;][]{Zeldovich:1967,Hawking:1971ei,car74}. PBHs would make up all or a part of dark matter (DM) \citep[see e.g.][for reviews]{car16}. Although Weakly Interacting Massive Particles (WIMPs) are considered as viable candidates \citep{jun96}, their existence has not been confirmed yet either by accelerator experiments, direct detection experiments, or indirect search experiments \citep{kla15,gas16}. 

The mass of PBHs surviving until present range from $10^{15}$g~\cite{Hawking:1971ei, car74} to as high as $10^5 M_\odot$~\cite{Kawasaki:2012kn}. To explore this wide range of masses,  various observational techniques and theoretical ideas have been employed~\cite{car16}. These include extragalactic gamma rays from PBH evaporation \citep{car10}, femtolensing of gamma-ray bursts \citep{bar12}, r-process nucleosynthesis \citep{ful17}, white-dwarf explosions \citep{gra15}, Subaru/HSC microlensing of stars in M31 \citep{nii17}, Kepler microlensing of stars \citep{gri14}, MACHO/EROS/OGLE mcirolensing of stars \citep{tis07} and quasar microlensing \citep{med09}, survival of a star cluster in Eridanus~H \citep{bra16}, wide binary disruption \citep{qui09}, millilensing of quasars \citep{wil01}, generation of large-scale structure through Poission fluctuations \citep{afs03}, dynamical friction on halo objects \citep{car99}, non-detection of gravitational wave background \citep{rai17}, and accretion effects on the cosmic microwave background \citep{ric08,ali17}. 

Recently, gravitational waves from binary black hole mergers have been detected by LIGO~\citep{abb16}. The origin of those binary black hole systems is under debate. PBHs are among the possible progenitors~\cite{Nakamura:1997sm,Clesse:2015wea,sas16,bir16,kaw16,inomata16}. Some published  constraints in the stellar and intermediate mass range have been recently questioned~\cite{bir16}, and there is a strong interest in finding additional reliable constraints at this range of masses. 

Freely-floating black holes (BHs), such as kicked stellar remnant BHs, are expected to interact with ambient interstellar medium (ISM) gas through Bondi-Hoyle-Lyttleton accretion \citep[e.g.][]{ips77,ips82,fuj98,ago02,mii05,iok16,mat17}. PBH in the stellar or intermediate mass ranges should also emit significant fluxes of X-ray photons through this process \cite{bar79,car79}. By comparing electromagnetic wave data with expected radiation signals from floating PBHs, one can set constraints on the abundance of PBHs \citep{bar79,car79,gag16}.

In this paper, we set new constraints on PBH abundance by utilizing the observed number density of X-ray binaries (XRBs) including ultra-luminous X-ray sources (ULXs). X-ray observations have already identified many X-ray emitting compact extragalactic objects, mostly XRBs. Moreover, we take into account a thick DM disk, so-called dark disk, which is suggested by numerical simulations \citep{rea08,rea09}. In Sec. \ref{sec:model}, we introduce accretion processes on to floating PBHs. Constraints on PBH abundances are presented in Sec. \ref{sec:result}. Discussion and Conclusions are given in Sec. \ref{sec:summary}.

\section{Accretion onto Floating PBHs}
\label{sec:model}

The accretion rate onto a floating PBH from the ISM is given by the Bondi-Hoyle-Lyttleton rate \citep{hoy39,bon44,bon52}: 
\begin{eqnarray}
\dot{M}&=&4\pi r_B^2 \tilde{v}\rho = \frac{4\pi G^2 M_{\rm BH}^2 n \, \mu \, m_p}{\tilde{v}^3}\\ \nonumber
&\simeq& 7.1\times10^{-2} \frac{L_{\rm Edd}}{c^2} \left(\frac{n}{100~{\rm cm^{-3}}}\right)\left(\frac{M_{\rm BH}}{100M_\odot}\right) \\
&\times&\left(\frac{\tilde{v}}{10~{\rm km~s^{-1}}}\right)^{-3},
\label{eq:mdot}
\end{eqnarray}
where $M_{\rm BH}$ is the PBH mass, $n$ is the ISM gas density, $m_p$ is the proton mass, $\mu$ is the mean molecular weight $\mu=2.72$, $L_{\rm Edd}$ is the Eddington luminosity, the Bondi radius $r_B=GM_{\rm BH}/\tilde{v}^2$, and $\tilde{v}\equiv(v^2+c_s^2)^{1/2}$. $v$ is the PBH velocity relative to the ISM gas cloud and $c_s$ is the sound speed which depends on the gas temperature. The effective value of the speed of sound for ISM gas is $c_s\sim10~{\rm km~s^{-1}}$ because ISM gas phases have a turbulent velocity of $\sim10~{\rm km~s^{-1}}$ in approximate pressure balance with each other. The mass accretion processes on to floating BHs is extensively investigated by \citep{ago02}. We basically follow their arguments hereafter.

The luminosity of a floating PBH interacting with ISM gas is estimated as
\begin{eqnarray}
\label{eq:L_Mdot}
L &=& \epsilon(\dot{M})\dot{M}c^2\\ \nonumber
&\simeq&4.7\times10^{37} {\rm erg~s^{-1}} \left(\frac{ \epsilon[\dot{M}]}{0.1}\right) \left(\frac{n}{100~{\rm cm^{-3}}}\right)\left(\frac{M_{\rm BH}}{100M_\odot}\right)^2\\
&\times&\left(\frac{\tilde{v}}{10~{\rm km~s^{-1}}}\right)^{-3},
\end{eqnarray}
where $\epsilon$ is the disk radiative efficiency. In the radiatively-inefficient accretion flow (RIAF) regime $L\lesssim0.01L_{\rm Edd}$, the luminosity is known to scale with $L\propto\dot{M}^2$ rather than $L\propto\dot{M}$ as in the standard accretion disk regime \citep{kat08}. Therefore, we set the radiative efficiency as
$\epsilon(\dot{M})=\epsilon_0/[1.0+({\dot{M}}/{0.01\dot{M}_{\rm Edd}})^{-1}]$, where $\dot{M}_{\rm Edd}\equiv L_{\rm Edd}/\epsilon_0 c^2$ is the Eddington mass accretion rate at the efficiency of $\epsilon_0=0.1$. We note that $\epsilon_0$ can be from 0.057 for a Schwarzschild BH to 0.42 for an extreme Kerr BH \citep[e.g.][]{kat08}.

Does gas form an accretion disk? Small perturbations in the density or velocity of the accreting gas lead to an angular momentum large enough to form a disk~\citep{sha76}. The accreted angular momentum is $l=(1/4)({\Delta \rho}/{\rho})\tilde{v}r_B$, where $\Delta\rho$ is the difference in density between the top and bottom of the accretion cylinder. The ISM density has turbulence with a Kolomogorov spectrum of fluctuations $\delta\rho/\rho\sim(L/10^{18}~{\rm cm})^{1/3}$ extending down to $\sim10^{8}~{\rm cm}$ \citep{arm95}. We can find the radius of the accretion disk $r_{\rm disk}$ by setting $\Delta\rho/\rho=\delta\rho/\rho|_{L=2r_B}$ and equating the angular momentum of the gas with the Keplerian angular momentum $l_K=\sqrt{GM_{\rm BH}r_{\rm disk}}$. This gives
\begin{eqnarray}
\frac{r_{\rm disk}}{r_s}&=&\frac{1}{16}\left(\frac{\delta\rho}{\rho}|_{L=2r_B}\right)^2\frac{r_B}{r_s}\\
&\simeq&2.5\times10^{6}\left(\frac{M_{\rm BH}}{100M_\odot}\right)^{2/3} \left(\frac{\tilde{v}}{10~{\rm km~s^{-1}}}\right)^{-10/3},
\end{eqnarray}
where $r_s$ is the Schwarzschild radius $2GM_{\rm BH}/c^2$. Thus, an accretion disk can be formed around a PBH by the Bondi-Hoyle-Lyttleton accretion.

We estimate the accretion timescale as $t_{\rm acc}\approx L_{\rm MC}/v \sim 2.4\times10^6(n/100~{\rm cm^{-3}})(v/10~{\rm km~s^{-1}})^{-1}~{\rm yr}$ which is longer than the observation periods of various X-ray sources since the dawn of X-ray astronomy. $L_{\rm MC}\sim24.7(n/100~{\rm cm}^{-3})^{-0.9}~{\rm pc}$ is the size of a typical molecular cloud~\cite{lar81}.

The number of PBHs interacting with the ISM gas can be expressed in terms of the luminosity function: 
\begin{equation}
\frac{dN}{dL_X}=\frac{dN}{d\dot{M}}\frac{d\dot{M}}{dL_X},
\end{equation}
where $dN/\dot{M}$ is the number of PBHs accreting at $\dot{M}$. Function $d\dot{M}/dL_X$ can be derived from Eq.(\ref{eq:L_Mdot}), and $dN/\dot{M}$ is given by \cite{ago02,mii05}
\begin{eqnarray}
\nonumber
\frac{dN}{d\dot{M}}&=&N_{\rm PBH, disk}\int_{n_{\rm min}}^{n_{\rm max}}dn\int_{0}^{\infty} dv \frac{df_n}{dn}\frac{df_v}{dv}\delta[\dot{M}(n,v)-\dot{M}]\\
&=&N_{\rm PBH, disk}\int_{n_{\rm min}}^{n_{\rm max}}dn\frac{df_n}{dn}\frac{df_v}{dv}(v_0)\frac{\tilde{v}^2_{v=v_0}}{3v_0\dot{M}}, \label{eq:dn_dMdot_fid}
\end{eqnarray}
where the number of PBHs passing through the Galactic disk region is given by $N_{\rm PBH, disk}$ and $v_0^2=(4\pi G^2M^2n\mu m_p/\dot{M})^{2/3}-c_s^2$. $df_n/dn$ and $df_v/dv$ is the probability distribution of the number density of the ISM gas and the velocity distribution of PBHs, respectively. We can not take all values of $n$ between $n_{\rm min}$ and $n_{\rm max}$ since $v_0^2>0$ and $n$ must be greater than $c_s^{3}/(4\pi G^2M^2\mu m_p/\dot{M})$.

Using the Navarro-Frenk-White (NFW) profile \citep{nav97} with parameters given in \citep{kly02} for the Milky way, the DM mass included within 15~kpc radius is about $\sim$7\% of $M_{\rm halo}=10^{12}M_\odot$. $N_{\rm PBH, disk}$ can be estimated as $0.07 (\pi r^2 2H_g)/(4\pi r^3/3)N_{\rm BH, tot}\sim0.11H_g/(15~{\rm kpc})N_{\rm BH, tot}$, where $N_{\rm PBH, tot}$ is the total number of PBHs in our galaxy $(\Omega_{\rm PBH}/{\Omega_M})(M_{\rm halo}/M_{\rm BH})$. $H_g$ is the disk scale height of the ISM gas. Recently, star formation activity in the extreme outer Galaxy at 22~kpc away from the Galactic center has been reported \citep{izu14}. Since the radial distribution of such clouds is not well understood yet, we do not take into account such distant clouds. If we  extrapolate the ISM gas distribution out to 22~kpc, we would expect a 10~\% increase in the X-ray emitting PBH number density.

The probability distribution of the number density of the ISM gas is defined as \citep{ago02,mii05}
\begin{equation}
\frac{df_n}{dn}=f_0 (n/n_{\rm min})^{-\beta},~(n_{\rm min}\leq n \leq n_{\rm max}).
\end{equation}
The normalization factor is defined as $f_0\equiv{(\beta-2)\langle \Sigma \rangle }/{2\mu m_p n_{\rm min}^2 H_g}$, where $\langle \Sigma \rangle$ is the mean surface mass density of the ISM gas in the Galactic disk, which is averaged for PBHs passing through the disk by $\langle \Sigma \rangle=\int2\pi r \Sigma(r)\rho_{\rm DM}(r)dr/\int2\pi r\rho_{\rm DM}(r)dr$, where $r$ is the Galactocentric radius and $\rho_{\rm DM}(r)$ is the dark halo density given by the NFW profile. 

For the ISM gas, we consider three gas phases, galactic disk molecular clouds, cold HI clouds, and the central molecular zone (CMZ). We do not take into account warm or hot ISM gas, which would host only low luminosity objects below X-ray detection thresholds. We take $\Sigma(r)$ of \citep{san84} and \citep{sco87}, for the molecular clouds and the HI clouds, respectively. We find $\langle \Sigma \rangle = 27 M_\odot~{\rm pc}^{-2}$ for the molecular clouds and $2.8 M_\odot~{\rm pc}^{-2}$ for the cold HI clouds.
The mass of CMZ is $\sim3\times10^7M_\odot$ and its radius is $\sim150~{\rm pc}$ with a scale height of $H_g=30~{\rm pc}$ \citep[see][and references therein]{tsu15}. In this paper, we assume a constant radial distribution of the surface mass density leading $\langle \Sigma \rangle = 420 M_\odot~{\rm pc}^{-2}$ for the CMZ region. Although the mass function of molecular cloud cores has been recently studied \citep{tsu15}, the density probability distribution function is not well constrained. Since the CMZ is composed of molecular clouds, we assume the same number density distribution function shape as that of molecular clouds. For the scale height of the molecular clouds and the HI clouds, we take $H_g=75$~pc and 150~pc, respectively, following \citep{ago02}. For the probability distribution function of the ISM gas, we take $\beta=2.8$, $n_{\rm min}=10^2~{\rm cm^{-3}}$, and $n_{\rm max}=10^5~{\rm cm^{-3}}$ for the molecular gas and the CMZ, while $\beta=3.8$, $n_{\rm min}=10~{\rm cm^{-3}}$, and $n_{\rm max}=10^2~{\rm cm^{-3}}$ for the HI atomic gas \citep{ago02}.

The velocity distribution of PBHs can be described by a Maxwell-Boltzmann distribution,
\begin{equation}
\frac{df_v}{dv}=\sqrt{\frac{2}{\pi}}\frac{v^2}{\sigma_v^3}\exp\left(-\frac{v^2}{2\sigma_v^2}\right).
\end{equation}
We take the velocity dispersion $\sigma_v=150~{\rm km~s^{-1}}$ \citep{lin10}, because PBHs are expected to make some fraction of DM particles.

Dynamical friction from gas reduces the speed of PBHs in molecular clouds \citep{ost99,mii05}. The dynamical friction force is given by \citep{ost99}
\begin{equation}
F_{\rm df}=-\frac{4\pi G^2 M_{\rm BH}^2n\mu m_p}{v^2}\ln\left(\frac{r_{\rm max}}{r_{\rm min}}\right),
\end{equation}
where $r_{\rm max}$ and $r_{\rm min}$ are the sizes of the surrounding medium and the object receiving the force, respectively. We take  $r_{\max}=L_{\rm MC}$ and $r_{\rm min}=r_s$. Then, by taking into account dynamical friction during the crossing time, one can rewrite Eq.(\ref{eq:dn_dMdot_fid}) as

\begin{equation}
\frac{dN}{d\dot{M}}= N_{\rm PBH, disk}\int_{n_{\rm min}}^{n_{\rm max}}dn\frac{df_n}{dn}\frac{df_v}{dv}(v_0)\frac{\tilde{v'}^2_{v=v_0}{v'_0}^2}{3v_0^3\dot{M}},
\end{equation}

where $\tilde{v'}^2=({v'_0}^2+c_2^2)^{1/2}$, ${v'_0}^2=(4\pi G^2M^2n\mu m_p/\dot{M})^{2/3}-c_s^2$, and $v_0^4 = {v'_0}^4 + \\16\pi G^2 M_{\rm BH}n\mu m_pL_{\rm MC}\ln(L_{\rm MC}/r_s)$. Hereafter, we take into account the effect of the dynamical friction.

Numerical simulations suggest that a thick DM disk, so-called dark disk, forms in a galactic halo after a merger at $z<2$ due to existence of stellar/gas disk \citep{rea08,rea09}. The dark disk corotates with the stellar disk with a scale height of $\sim1.5~{\rm kpc}$. The dark disk can contribute to $f_{\rm DM, disk}\sim0.25$-1.5 times of the non-rotating halo DM density at the solar position. The lower bound is the conservative limit since it takes into account only one merger with a Large Magellanic Cloud mass. The velocity dispersion of the dark disk is $50~{\rm km~s^{-1}}$ \citep{rea08,rea09,bru09,lin10b}. This low velocity would boost up interaction rates. We assume a radially constant $f_{\rm DM, disk}$ in the disk, which is reasonable assumption for our order-of-magnitude calculation \citep{rea09}.

The luminosity functions of XRBs have been established in literature and they are also known to be correlated with star formation rate (SFR) of their host galaxies \citep[e.g.][]{gri03,min12}. We adopt the SFR normalized luminosity function in \citep{min12}, where the SFR is determined for the stellar mass range of $0.1-100~M_\odot$ with a Salpeter initial mass function. We take the SFR of the Milky way as $\sim1~M_\odot~{\rm yr}^{-1}$ \citep[e.g.,][]{rob10}.

\begin{figure*}[t]
\includegraphics[width=12.0cm]{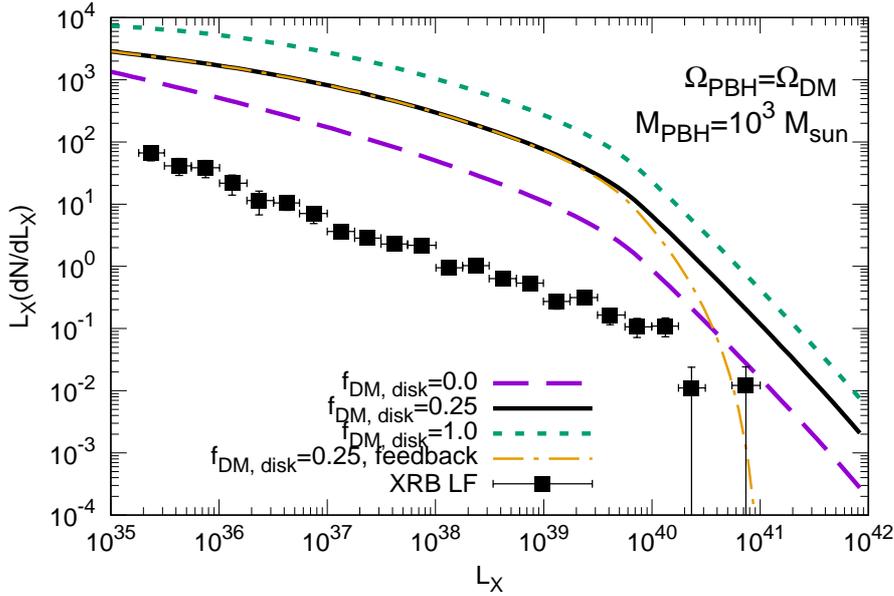}% Here is how to import EPS art
\caption{\label{fig:XLF} X-ray luminosity function of PBHs passing through ISM gas for $M_{\rm BH}=10^3M_\odot$ and $\Omega_{\rm PBH}=\Omega_{\rm DM}$. Dashed, solid, and dotted curve corresponds to $f_{\rm DM, disk}=0.0$, 0.25, and 1.0 accounting for all the ISM target gases. Dot-dashed curve takes into account radiation feedback for the case of $f_{\rm DM, disk}=0.25$. The values are normalized to SFR$=1~M_\odot~{\rm yr}^{-1}$. The data points show the observed SFR normalized X-ray luminosity function from \citep{min12}.}
\end{figure*}

\section{Results}
\label{sec:result}

The results are shown in Fig. \ref{fig:XLF} for $M_{\rm BH}=10^3M_\odot$ assuming $\Omega_{\rm PBH} = \Omega_{\rm DM}$. It is clear that $\Omega_{\rm PBH}$ need to be less than $\Omega_{\rm DM}$ even for $f_{\rm DM, disk}=0.0$, otherwise it will violate observations of XRBs. By taking into account the dark disk component, the luminosity function is enhanced because of its low velocity dispersion. Interaction with molecular clouds dominate the luminosity function. Other ISM components will not contribute significantly. 

The PBH density is constrained by data points having X-ray luminosities above $10^{39}~{\rm erg~s^{-1}}$ corresponding to ULXs. Those luminous objects would prevent efficient mass accretion as given in Equation \ref{eq:mdot} due to radiation feedback effects. Mass accretion rate is known to decrease by a factor of $(1-L/L_{\rm Edd})^2$ in the Bondi-Hoyle-Lyttleton accretion \citep{fuk99}. The radiation feedback effect is also shown in Figure. \ref{fig:XLF}. Although it will decrease the XLF at high luminosity end, the constraint will not be significantly different. 

\begin{figure*}[t]
\includegraphics[width=12cm]{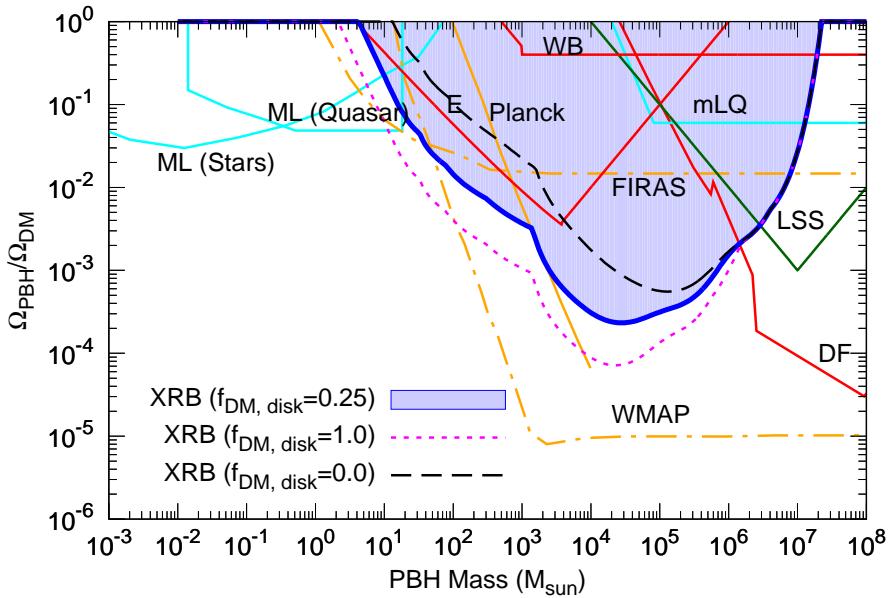}% Here is how to import EPS art
\caption{\label{fig:Omega} The upper bound on the PBH mass fraction relative to DM based on luminosity function of XRBs. The shaded region corresponds to the fiducial case with $f_{\rm DM, disk}=0.25$. Dashed and dotted line corresponds to the case with $f_{\rm DM, disk}=0.0$ and $1.0$, respectively. Other constraints are MACHO/EROS/OGLE mcirolensing of stars (ML) \citep{tis07} and quasar microlensing (ML) \citep{med09}, survival of a star cluster in Eridanus~H (E) \citep{bra16}, wide binary disruption (WB) \citep{qui09}, millilensing of quasars (mLQ) \citep{wil01}, generation of large-scale structure through Poission fluctuations (LSS) \citep{afs03,car10}, dynamical friction on halo objects (DF) \citep{car99}, and accretion effects on the cosmic microwave background  using the FIRAS data (FIRAS) and the WMAP data (WMAP) \citep{ric08} and the Planck data (Planck) \citep{ali17}. Since the constraints from the CMB data have been recently revisited by the Planck data, we show the previous CMB constraints in dot-dashed lines. The conservative limit is shown for the Planck data. We do not show the constraint from the Planck data above $10^4M_\odot$ because the constraint above this mass is not shown in the original reference.}
\end{figure*}

Constraints on $f_{\rm PBH}\equiv\Omega_{\rm PBH}/\Omega_{\rm DM}$ can be set based on the requirement that the predicted luminosity function does not violate the observed luminosity function of XRBs at any luminosities. In Fig.~\ref{fig:Omega}, we show the upper bound on the PBH mass fraction to DM based on the XRB luminosity function together with other constraints. For simplicity, we assume a monochromatic mass distribution for PBHs. We show three limits. One is our fiducial model with $f_{\rm DM, disk}=0.25$. The others are the cases with $f_{\rm DM, disk}=0.0$ and $1.0$. 

If we do not take into account the effect of dynamical friction, the constraints can become  $\sim3$\% tighter at $10~M_\odot\lesssim M_{\rm PBH}\lesssim1000~M_\odot$. This is because dynamical friction changes the velocity by about 1~${\rm km~s}^{-1}$ in our parameter space.

For the ISM gas phases, we assumed $n_{\rm min}=10^2~{\rm cm^{-3}}$ for the molecular gas and the CMZ and $n_{\rm min}=10~{\rm cm^{-3}}$ for the HI atomic gas \citep{ago02}. Given in Eq. \ref{eq:L_Mdot}, luminosities of accreting massive PBHs become too high for observed XRB luminosities (See. Fig. \ref{fig:XLF}). Thus, we see an upper cutoff at high PBH mass range in our constraints.

\section{Discussion and Conclusions}
\label{sec:summary}

Recently, the constraint by \citep{ric08} utilizing the cosmic microwave background (CMB) data is revisited by \citep{ali17} with the latest Planck data by taking account into detailed physical processes. For example, the radiative efficiency is estimated from the first principle in \citep{ali17}, while it is fixed in \citep{ric08}. As a consequence, the CMB constraints on the PBH density is significantly weakened (Fig.~\ref{fig:Omega}). However, we note that \citep{che16} obtained a tighter constraint than that by \citep{ric08} with Planck data following the method in \citep{ric08}.

Our methods constrain the PBH abundance in the mass range from a few $ M_\odot$ to $2\times 10^7 M_\odot$. The obtained constraint is tighter than the other constraints at $10~M_\odot\lesssim M_{\rm PBH}\lesssim1000~M_\odot$. Since we consider the ISM gas phases of molecular clouds, HI clouds, and CMZ whose density is at $10~{\rm cm^{-3}}\leq n \leq 10^{5}~{\rm cm^{-3}}$, we can not put tight constraints at low and high mass end of PBHs with X-ray data ranging $35\leq \log L_X \leq 41$. 

Very recently, \citep{gag16} has also studied the PBH abundances based on the study of Bondi accretion processes in Galactic PBHs. Although our estimates are based on different reasoning, the results are consistent for the case of $f_{\rm DM, disk}=0.0$. The approach of Ref.~\cite{gag16} is based on comparison with the X-ray and radio source catalogs in the Galactic ridge region, while we use the luminosity function of x-ray binaries. While in Ref.~\cite{gag16} it is assumed that all the sources listed in the X-ray catalogs are BH candidates, it is known that white dwarfs and coronally active stars are the dominant sources in the Galactic ridge region~\cite{rev09}. To make use of the radio catalog data  (which we do not use) the authors of Ref.~\cite{gag16} converted x-ray luminosities to radio luminosities using the fundamental plane of black holes~\citep{mer03}. However, the existence of the plane is questioned by recent studies \citep{ino17}. 

At $30 M_\odot$, we find $f_{\rm PBH}\lesssim4\times10^{-2}$ with our fiducial model $f_{\rm DM, disk} = 0.25$. This bound rules out PBH scenarios requiring $f_{\rm PBH}$ of an order of unity to explain the GW events \citep{bir16}, while our limit still marginally allows the models in which binary PBH systems are efficiently formed in the radiation dominant era \citep{iok98} requiring $f_{\rm PBH}\sim2\times10^{-3}-2\times10^{-2}$ at $30 M_\odot$ \citep{sas16,ero16,hay16}. Further detailed studies of $f_{\rm DM,disk}$ and the binary black hole merger rate will tighten the current limit on to the PBH scenario as the dominant GW sources.

X-ray emission from free-floating BHs interacting with ISM gas is similar to XRBs~\citep{ips77,ips82,fuj98,ago02,mii05,iok16,mat17}, but some features may help distinguish such isolated systems from XRB systems. First, x-ray emitting isolated BHs must be associated with molecular clouds or cold HI clouds if they are to emit significant fluxes of x-ray photons (Eq. \ref{eq:L_Mdot}). Second, one will not see any orbital motion, because the BHs are isolated. Lastly, these isolated BHs are expected to be associated with ionized bubbles, because the ISM gas needed for the emission is ionized by the emitted x-rays. Thus, a spatially extended iron K$\alpha$ fluorescent line structure can be found, although one needs to resolve the source of the size $L_{\rm MC}$.
%The expected equivalent width can be estimated as \citep[See Eq. 4 in][]{tsu05} ${\rm EW}_{\rm K \alpha}\simeq 1.9(n/100~{\rm cm}^{-3})^{0.1}~{\rm eV}$. 

To summarize, we have set a new observational bound on the abundance of PBHs in the mass range from a few $ M_\odot$ to $2\times 10^7 M_\odot$ utilizing X-ray data by considering PBH interactions with interstellar medium. The density of ULXs gives the most stringent constraint on the PBH abundance. Our constraints are tighter than recent constraints by Planck data at $5M_\odot\lesssim M_{\rm PBH}\lesssim300M_\odot$. 

\acknowledgments
The authors thank Yutaka Fujita, Kunihito Ioka, Norita Kawanaka, Tomonori Totani, Masato Tsuboi, Masahiro Tsujimoto for comments and discussions. YI is supported by the JAXA international top young fellowship and JSPS KAKENHI Grant Number JP16K13813.  The work of A.K. was supported by the U.S. Department of Energy Grant No. DE-SC0009937, and by the World Premier International Research Center Initiative (WPI), MEXT, Japan. 

\bibliographystyle{JHEP}
\bibliography{pbhbib}% Produces the bibliography via BibTeX.
\end{document}